\documentclass[12pt]{iopart}

\usepackage{graphicx}
\usepackage{iopams,latexsym}
\usepackage{amssymb}
\usepackage{epsf}
\usepackage{color}

\begin{document}

\title[Interplay of degeneracy and non-degeneracy .......... feed-forward loop motif]{Interplay of degeneracy and non-degeneracy in fluctuations propagation in coherent feed-forward loop motif}

\author{Tuhin Subhra Roy$^1$, Mintu Nandi$^2$, Pinaki Chaudhury$^3$,
Sudip Chattopadhyay$^2$, and Suman K Banik$^1$}

\address{$^1$Department of Chemistry, Bose Institute, 93/1 A P C Road, Kolkata 700009, India}
\address{$^2$Department of Chemistry, Indian Institute of Engineering Science and Technology, Shibpur, Howrah 711103, India}
\address{$^3$Department of Chemistry, University of Calcutta, 92 A P C Road, Kolkata 700009, India}

\ead{tsroy@jcbose.ac.in, mintu.rs2022@chem.iiests.ac.in, pcchem@caluniv.ac.in, sudip@chem.iiests.ac.in, skbanik@jcbose.ac.in}

\date{\today}

\begin{abstract}
We present a stochastic framework to decipher fluctuations propagation in classes of coherent feed-forward loops. The systematic contribution of the direct (one-step) and indirect (two-step) pathways is considered to quantify fluctuations of the output node. We also consider both additive and multiplicative integration mechanisms of the two parallel pathways (one-step and two-step). Analytical expression of the output node's coefficient of variation shows contributions of intrinsic, one-step, two-step, and cross-interaction in closed form. We observe a diverse range of degeneracy and non-degeneracy in each of the decomposed fluctuations term and their contribution to the overall output fluctuations of each coherent feed-forward loop motif. Analysis of output fluctuations reveals a maximal level of fluctuations of the coherent feed-forward loop motif of type 1.
\end{abstract}

\maketitle


\section{Introduction}

Through the course of evolution, living systems developed complex biochemical networks using multiple nodes and edges. The gene transcription regulatory networks (GTRNs) are one such network where the nodes are represented by genes and the edges take care of the directed interaction of two adjacent genes \cite{Milo2002,Alon2006}. Depending on the type of simple regulation (positive/negative) the edges could be activating or repressing in nature \cite{Alon2006}. A crucial function of a GTRN is to regulate the propagation of cellular fluctuations that arise due to the probabilistic nature of interaction \cite{Kaern2005,Eldar2010}.


\begin{figure}[!t]
\includegraphics[width=0.75\columnwidth,angle=90]{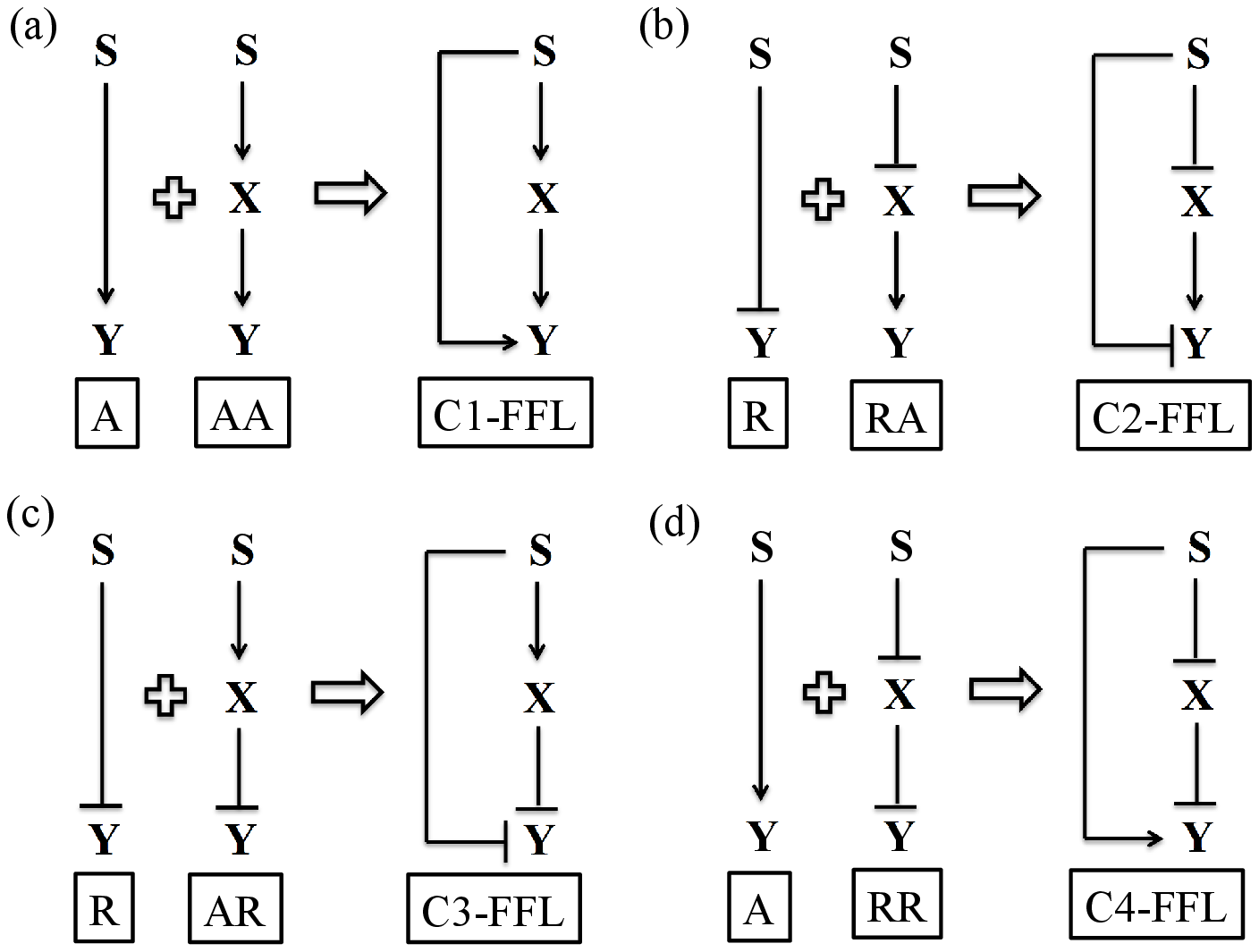}
\caption{Schematic diagram of in built components of CFFL. A and R represent pure one-step cascade motifs of activating and repressing in nature, respectively. AA, AR, RA, and RR represent pure TSC motifs.  
(a)  C1-FFL is made of A (S~$\rightarrow$~Y) and AA (S~$\rightarrow$~X~$\rightarrow$~Y).
(b)  C2-FFL is made of R (S~$\dashv$~Y) and RA (S~$\dashv$~X~$\rightarrow$~Y).
(c)  C3-FFL is made of R (S~$\dashv$~Y) and AR (S~$\rightarrow$~X~$\dashv$~Y).
(d)  C4-FFL is made of A (S~$\rightarrow$~Y) and RR (S~$\dashv$~X~$\dashv$~Y).
In the diagram, $\rightarrow$ and $\dashv$ stand for activation and repression, respectively.
In CFFL, integration of fluctuations into the target Y takes place using additive or multiplicative mechanism (see text).
}
\label{fig1}
\end{figure}

One of the well studied GTRN is the feed-forward loop (FFL) network \cite{Shen-Orr2002,Mangan2003}. In a generic FFL, the final output is regulated by two parallel pathways, viz., direct pathway and indirect pathway \cite{Alon2006}. In a direct path, two nodes are connected by a single directed edge, thus mimicking a one-step cascade (OSC). On the other hand, an indirect path has three nodes connected by two directed edges which resembles a two-step cascade (TSC). Depending on the nature of the edges (activating/repressing) the overall sign of the indirect pathway may or may not be equal to the sign of the direct pathway. When the overall sign of the indirect pathway is the same as the sign of the direct pathway one ends up with a coherent feed-forward loop (CFFL) motif. On the other hand, in an incoherent feed-forward loop (IFFL) motif, the overall sign of the indirect pathway is different from the sign of the direct pathway. Furthermore, each of the two broad categories is subdivided into four classes thus providing total 8 types of FFL motifs. Out of the 8 types of FFL motifs only two types, viz., C1-FFL and I1-FFL are more frequent as observed in \textit{Escherichia coli} network \cite{Shen-Orr2002,Mangan2003,Ma2004,Mangan2006}. Several experimental works on FFL network have been reported over the years \cite{Mangan2003,Kalir2005,Mangan2006,Amit2007,Kaplan2008,Bleris2011,Chen2013,Sasse2015}. CFFLs are abundant in organisms including \textit{Escherichia coli}, yeast, Drosophila nervous system, and several higher organisms \cite{Alon2006,Alon2007,Li2009,Herranz2010}. Mangan and Alon \cite{Mangan2003} studied the deterministic dynamics of FFLs. The fluctuations propagation and its relation to the architecture of the CFFLs have been studied earlier \cite{Ghosh2005,Kittisopikul2010,Maity2015,Momin2020a,Biswas2022}. CFFLs were found to have efficient fluctuations propagation property along the cascade \cite{Dunlop2008}.
Theoretical analysis of FFL network within the context of cellular fluctuations has been also addressed \cite{Ghosh2005,Maity2015,Momin2020a,Momin2020b,Nandi2021,Biswas2022}. The role of timescale separation of the network components in fluctuations propagation was addressed earlier by Maity et al \cite{Maity2015}. Theoretic analysis reveals C1-FFL as an efficient information transmission motif \cite{Momin2020a}. A pathway-based decomposition of the output fluctuations addressed the redundant information processing capacity of C1- and I1-FFL \cite{Biswas2022}.

In the present work, we aim to decipher fluctuations propagation in classes of CFFL networks. Depending on the nature of edges (positive/negative) in a CFFL network, the effective sign of the indirect path is the same as the sign of the direct path. In a CFFL, one can consider the indirect path and the direct path as two-step and one-step pathways, respectively (see Fig.~\ref{fig1}). A pure TSC is a simple and recurring building block of the GTRN and often occurs in several bacterial organisms \cite{Alon2006}. A TSC consists of three regulatory molecules S, X, and Y (say), where S regulates X which in turn regulates Y. The regulations can be of activating or repressing in nature. Different combinations of activation and repression edges in a TSC motif lead to four types of structures, viz., activating-activating (AA), repressing-activating (RA), activating-repressing (AR), and repressing-repressing (RR) (see Fig.~\ref{fig1}).

In response to a common signal, genes of a gene regulatory network develops the pool of gene products of which some act as transcription factors for downstream genes. Depending on the architecture of the network and nature of regulation (i.e., activation and repression) some of the gene product's fluctuations levels are exactly similar giving rise to collating profiles in response to the common signal. The collating or degenerate profiles help one to classify the cascades or motifs into different categories. In many cases, however, one observes a non-degenerate response.
In a recent study, \cite{Roy2021}, AA and AR motifs are found to collate with respect to the fidelity of signal transduction (measured as signal-to-noise ratio (SNR)). In addition, RA and RR motifs are also found to have a collating nature with respect to the said metric.
Such a degeneracy among the TSC motifs characterizes their equal ability in signal propagation mechanism as reflected through mutual information and fidelity. A signature of degeneracy can be useful as it reveals indistinguishable responses, of classes of GTRN and its building blocks, to a common signal at a steady state. For a complex GTRN, such as FFL, degeneracy might help in showing similarity in response of the in-built sub-structures, e.g., OSC and TSC, and their cumulative contribution in fluctuations propagation.



\begin{figure}[!t]
\includegraphics[width=0.75\columnwidth,angle=-90]{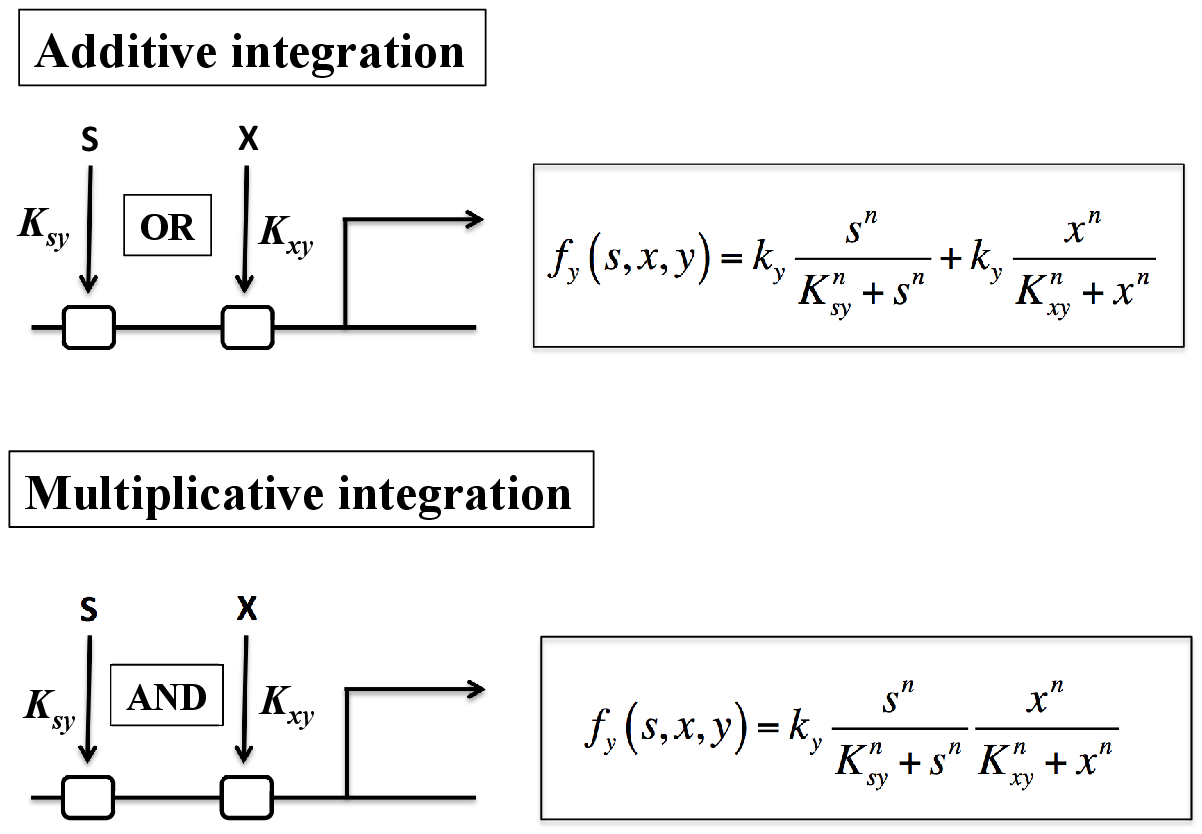}
\caption{Schematic presentation of the interaction of S and X with the two binding sites (open rectangles) of the promoter of Y. Additive and multiplicative integration mechanism for C1-FFL is shown along with mathematical expression used in the paper.
}
\label{fig2}
\end{figure}

Here, we integrate an OSC (of activating (A) or repressing (R) nature) laterally with these TSC motifs. 
The lateral integration is done in such a way that it leads to four types of CFFLs (see Fig.~\ref{fig1}). We note that, among these CFFLs, C1-FFL is most abundant in bacterial GTRN \cite{Alon2006}. 
Signal propagation through the parallel pathways (direct and indirect) gets integrated into the final output. Two different integration mechanisms are considered keeping in mind the biological aspects of gene regulation \cite{Alon2006}. In one case, one of the inputs (transcription factors) from the parallel pathways is enough to regulate the output gene expression thus leading to an additive integration mechanism (see Fig.~\ref{fig2}). When both the inputs (transcription factors) from the parallel pathways are necessary for the regulation of the output gene, the system follows the multiplicative integration mechanism (see Fig.~\ref{fig2}). We note that, in the multiplicative integration mechanism, the level of both inputs needs to cross a threshold. For an additive integration mechanism, however, the level of one of the two inputs needs to achieve the necessary threshold. Achievement of the threshold of a particular input (transcription factor) is taken care of by the binding constant $K$. For specific binding constants, we refer to Table~\ref{table1}. The population of input (transcription factor) keeps track of the necessary threshold required for target gene regulation.
Previous communication \cite{Nandi2021} reveals the integration nature of the two pathways in FFLs is competitive for additive integration mechanism and non-competitive for multiplicative integration mechanism. Since one of the two inputs coming from parallel (direct and indirect) pathways is enough to regulate the target gene expression in the additive integration mechanism, a competition between the inputs is in action and depends on the necessary binding constant $K$. In the multiplicative integration mechanism, however, such competition disappears as both inputs are necessary for target gene regulation.
Theoretical analysis reveals an additional fluctuations term that appears due to the integration of OSC and TSC \cite{Momin2020a,Nandi2021,Biswas2022}. The additional fluctuations (or cross-interaction) originate due to the lateral overlap of the two parallel pathways (direct and indirect) as shown in Fig.~\ref{fig1}. A proper decomposition mechanism of the output's fluctuations yields the contribution of the direct, indirect, and cross-interaction in a CFFL.

In the present manuscript, we quantify the output fluctuations by measuring the coefficient of variation (CV) of output Y. To understand the nature of fluctuations propagation in CFFLs, we opt for the strategy of fluctuations decomposition reported earlier \cite{Paulsson2004,Swain2002,Momin2020a,Nandi2021,Biswas2022}. 
Fluctuations decomposition has been developed to quantify the gene expression mechanism in noisy environment \cite{Paulsson2004}. The formalism has also been applied to decipher the extrinsic and intrinsic contribution of noise processes in stochastic gene expression \cite{Swain2002}. 
We note that formalism of fluctuations decomposition helps in understanding the nature of fluctuations propagation in CFFLs with respect to pure TSC motifs (AA, RA, AR, and RR), where collating nature has been observed \cite{Roy2021}. In the next section, we discuss the stochastic model of CFFLs and corresponding mathematical analysis to compute fluctuations associated with output Y. In Sec.~\ref{sec3}, we analyze the results using the notion of fluctuations decomposition. Fluctuations decomposition characterizes degenerate and non-degenerate fluctuations propagations in components of a CFFL, viz., one-step pathway and two-step pathway. It also quantifies the contribution of cross-interaction terms appearing due to additive and multiplicative integration mechanisms. The paper is concluded in Sec.~\ref{sec4}.


\section{Model and Methods}
\label{sec2}

The stochastic dynamics of a generic FFL are can be expressed in terms of coupled Langevin equations,

\begin{eqnarray}
\label{eq1}
\frac{ds}{dt} & = & f_{s}(s)-\mu_s s+\xi_s(t), \\
\label{eq2}
\frac{dx}{dt} & = & f_{x}(s,x)-\mu_x x+\xi_x(t), \\
\label{eq3}
\frac{dy}{dt} & = & f_{y}(s,x,y)-\mu_y y+\xi_y(t).
\end{eqnarray}

\noindent In Eqs.~(\ref{eq1}-\ref{eq3}), $s, x$ and $y$ are the copy numbers of species S, X, and Y, respectively, expressed in molecules/$V$, where $V$ is the unit effective cellular volume. Here copy number stands for the population of a species. $f_{s}(s)$, $f_{x}(s, x)$ and $f_{y}(s, x, y)$ are the production term associated with $s$, $x$ and $y$, respectively. We note that the production term of each component, in general, are nonlinear in nature \cite{Bintu2005,Ziv2007,Tkacik2008a,Tkacik2008b,Ronde2012}. The degradation rate constants associated with $s$, $x$ and $y$ are $\mu_s$, $\mu_x$, and $\mu_y$, respectively. The constants $\mu_s$, $\mu_x$, and $\mu_y$ correspond to the inverse lifetime of the gene product of the respective component.

The noise processes $\xi_i (t)$ considered here are independent and Gaussian distributed with properties $\langle \xi_{i}(t) \rangle$ = 0 and $\langle \xi_{i}(t)\xi_{j}(t^{'}) \rangle$ = $\langle |\xi_{i}|^{2}\rangle \delta_{ij} \delta(t-t^{'})$ with $\langle |\xi_{i}|^{2}\rangle = \langle f_{i} \rangle + \mu_{i}\langle i \rangle = 2\mu_{i}\langle i \rangle$ ($i =s, x$, and $y$) at steady state \cite{Elf2003,Swain2004,Paulsson2004,Tanase2006,Warren2006,Kampen2007,Mehta2008,Ronde2010}. The nature of noise correlation also suggests that the noise processes are uncorrelated with each other \cite{Swain2016}. Here, $\langle \cdots \rangle$ represents steady state ensemble average. The noise processes we consider here are intrinsic in nature. As a result, variability induced in the dynamics is controlled by the intrinsic noise processes although it has been observed that fluctuations in gene expression in a genetically identical population are mostly controlled by extrinsic noise processes \cite{Elowitz2002,Raser2004}. Depending on the nature of interactions, activation or repression shown in Fig.~\ref{fig1}, we consider specific form of the functions $f_s (s)$, $f_x (s, x)$ and $f_y (s, x, y)$ (see Table~\ref{table1}).


\begin{table}[!t]
\small
\caption{List of functions used in Eqs.~(\ref{eq1}-\ref{eq3}) using different signal integration mechanism.}
\label{table1}
\begin{tabular}{lccc}
\br
Motif & $f_s(s)$ & $f_x(s,x)$ & $f_y(s,x,y)$ \\
\mr
C1-FFL (additive)  & $k_s$ & $k_x \frac{s^n}{K_{sx}^n + s^n}$ & 
$k_{y} \frac{x^n}{K_{xy}^n + x^n} + k_{y} \frac{s^n}{K_{sy}^n + s^n}$ \\
\\
C1-FFL (multiplicative)  & $k_s$ & $k_x \frac{s^n}{K_{sx}^n + s^n}$ & 
$k_{y} \frac{x^n}{K_{xy}^n + x^n} \frac{s^n}{K_{sy}^n + s^n}$ \\
\\
C2-FFL (additive) & $k_s$ & $k_x \frac{K_{sx}^n}{K_{sx}^n + s^n}$ & 
$k_{y} \frac{x^n}{K_{xy}^n + x^n} + k_{y} \frac{K_{sy}^n}{K_{sy}^n + s^n}$ \\
\\
C2-FFL (multiplicative) & $k_s$ & $k_x \frac{K_{sx}^n}{K_{sx}^n + s^n}$ & 
$k_y \frac{x^n}{K_{xy}^n + x^n} \frac{K_{sy}^n}{K_{sy}^n + s^n}$ \\
\\
C3-FFL (additive) & $k_s$ & $k_x \frac{s^n}{K_{sx}^n + s^n}$ & 
$k_{y} \frac{K_{xy}^n}{K_{xy}^n + x^n} + k_{y} \frac{K_{sy}^n}{K_{sy}^n + s^n}$ \\
\\
C3-FFL (multiplicative) & $k_s$ & $k_x \frac{s^n}{K_{sx}^n + s^n}$ & 
$k_y \frac{K_{xy}^n}{K_{xy}^n + x^n} \frac{K_{sy}^n}{K_{sy}^n + s^n}$ \\
\\
C4-FFL (additive) & $k_s$ & $k_x \frac{K_{sx}^n}{K_{sx}^n + s^n}$ & 
$k_{y} \frac{K_{xy}^n}{K_{xy}^n + x^n} + k_{y} \frac{s^n}{K_{sy}^n + s^n}$ \\
\\
C4-FFL (multiplicative) & $k_s$ & $k_x \frac{K_{sx}^n}{K_{sx}^n + s^n}$ & 
$k_y \frac{K_{xy}^n}{K_{xy}^n + x^n} \frac{s^n}{K_{sy}^n + s^n}$\\
\br
\end{tabular}
\end{table}

To calculate the second moments associated with $s, x$ and, $y$ we consider linearized kinetics \cite{Kampen2007} around steady state. A perturbation of linear order $\delta W(t) = W(t) - \langle W \rangle$ where $\langle W \rangle$ is the average population of $W$ at steady is used to recast Eqs.~(\ref{eq1}-\ref{eq3}) in the following form
\begin{equation}
\label{eq4}
\frac{d\mathbf{\delta W}}{dt} = \mathbf{J}_{W=\langle W \rangle}
\mathbf{\delta W}(t) + \mathbf{\Xi}(t).
\end{equation}

\noindent In the above equation, $\mathbf{\delta W}(t)$ is the fluctuations matrix containing the linear order perturbation terms. The noise matrix $\mathbf{ \Xi}(t)$ along with $\mathbf{\delta W}(t)$ can be written as
\begin{eqnarray*}
\mathbf{\delta W}(t) = \left( 
\begin{array}{ccc} \delta s(t) \\ \delta x(t) \\ \delta y(t) \\
\end{array} \right),
\mathbf{ \Xi}(t) = \left( 
\begin{array}{ccc}  \xi_{s}(t) \\  \xi_{x}(t) \\  \xi_{y}(t) \\
\end{array} \right).
\end{eqnarray*}

\noindent
$\mathbf{J}$ represents the Jacobian matrix at the steady state,
\begin{eqnarray*}
\mathbf{J} = \left (
\begin{array}{ccc}
f^{\prime}_{s,s} (\langle s \rangle) - \mu_s & 0 & 0 \\
f^{\prime}_{x,s} (\langle s \rangle, \langle x \rangle) &
f^{\prime}_{x,x} (\langle s \rangle, \langle x \rangle) - \mu_x & 0 \\
f^{\prime}_{y,s} (\langle s \rangle, \langle x \rangle, \langle y \rangle) &
f^{\prime}_{y,x} (\langle s \rangle, \langle x \rangle, \langle y \rangle) &
f^{\prime}_{y,y} (\langle s \rangle, \langle x \rangle, \langle y \rangle) - \mu_y
\end{array}
\right ).
\end{eqnarray*}

\noindent Here $f^{\prime}_{s,s} (\langle s \rangle)$ stands for differentiation of $f_s (s)$ with respect to $s$ and evaluated at $s = \langle s \rangle$, and so on. In the following, we will write $f^{\prime}_{s,s} (\langle s \rangle)$, $f^{\prime}_{x,s} (\langle s \rangle, \langle x \rangle)$, etc as $f^{\prime}_{s,s}$, $f^{\prime}_{x,s}$, etc, respectively, for notational simplicity. The variance and covariance is evaluated using the Lyapunov equation at steady state \cite{Keizer1987,Elf2003,Paulsson2004,Paulsson2005,Kampen2007}
\begin{equation}
\label{eq5}
\mathbf{J \Sigma} + \mathbf{\Sigma J}^{T} + \mathbf{D} = \mathbf{0}.
\end{equation}


\begin{table}[!t]
\small
\caption{List of $k_s$, $k_x$ and $k_y$ for different CFFLs using additive and multiplicative integration mechanism.}
\label{table2}
\begin{tabular}{lccc}
\br
Motif & $k_s$ & $k_x$ & $k_y$ \\
\mr
C1-FFL (additive)  & $\mu_s \langle s \rangle$ & $\mu_x \langle x \rangle \frac{K_{sx}^n + \langle s \rangle^n}{\langle s \rangle^n}$ & 
$\frac{\mu_y \langle y\rangle(K_{sy}^n+\langle s\rangle^n)(K_{xy}^n+\langle x\rangle^n)}{2\langle s\rangle^n\langle x\rangle^n+K_{sy}^n\langle x\rangle^n+K_{xy}^n \langle s\rangle^n} $ \\
\\
C1-FFL (multiplicative)  & $\mu_s \langle s \rangle$ & $\mu_x \langle x \rangle \frac{K_{sx}^n + \langle s \rangle^n}{\langle s \rangle^n}$  & 
$ \frac{\mu_y \langle y\rangle(K_{xy}^n+\langle x\rangle^n)(K_{sy}^n+\langle s\rangle^n)}{\langle s\rangle^n \langle x\rangle^n } $ \\
\\
C2-FFL (additive) & $\mu_s \langle s \rangle$ & $\mu_x \langle x \rangle \frac{K_{sx}^n + \langle s \rangle^n}{K_{sx}^n}$  & 
$\frac{\mu_y \langle y\rangle(K_{xy}^n+\langle x\rangle^n)(K_{sy}^n+\langle s\rangle^n)}{\langle s\rangle^n\langle x\rangle^n+2K_{sy}^n\langle x\rangle^n+K_{sy}^nk_{xy}^n}$ \\
\\
C2-FFL (multiplicative) & $\mu_s \langle s \rangle$ & $\mu_x \langle x \rangle \frac{K_{sx}^n + \langle s \rangle^n}{K_{sx}^n}$  & 
$  \frac{\mu_y \langle y\rangle(K_{xy}^n+\langle x\rangle^n)(K_{sy}^n+\langle s\rangle^n)}{K_{sy}^n\langle x\rangle^n} $ \\
\\
C3-FFL (additive) & $\mu_s \langle s \rangle$ & $\mu_x \langle x \rangle \frac{K_{sx}^n + \langle s \rangle^n}{\langle s \rangle^n}$ & 
$ \frac{\mu_y \langle y\rangle(K_{xy}^n+\langle x\rangle^n)(K_{sy}^n+\langle s\rangle^n)}{K_{xy}^n\langle s\rangle^n+K_{sy}^n\langle x\rangle^n+2K_{xy}^nk_{sy}^n} $ \\
\\
C3-FFL (multiplicative) & $\mu_s \langle s \rangle$ & $\mu_x \langle x \rangle \frac{K_{sx}^n + \langle s \rangle^n}{\langle s \rangle^n}$ & 
$  \frac{\mu_y \langle y\rangle(K_{xy}^n+\langle x\rangle^n)(K_{sy}^n+\langle s\rangle^n)}{K_{xy}^nk_{sy}^n} $ \\
\\
C4-FFL (additive) & $\mu_s \langle s \rangle$ & $\mu_x \langle x \rangle \frac{K_{sx}^n + \langle s \rangle^n}{K_{sx}^n}$ & 
$\frac{\mu_y \langle y\rangle(K_{xy}^n+\langle x\rangle^n)(K_{sy}^n+\langle s\rangle^n)}{\langle s\rangle^n\langle x\rangle^n+2K_{xy}^n\langle s\rangle^n+K_{xy}^nk_{sy}^n}  $ \\
\\
C4-FFL (multiplicative) & $\mu_s \langle s \rangle$ & $\mu_x \langle x \rangle \frac{K_{sx}^n + \langle s \rangle^n}{K_{sx}^n}$ & 
$  \frac{\mu_y \langle y\rangle(K_{xy}^n+\langle x\rangle^n)(K_{sy}^n+\langle s\rangle^n)}{K_{xy}^n\langle s\rangle^n} $\\
\br
\end{tabular}
\end{table}

\noindent In the Lyapunov equation, $\mathbf{\Sigma}$ and $\mathbf{D}$ stand for covariance matrix and diffusion matrix, respectively. The diffusion matrix $\mathbf{D}$ incorporates different noise strength using the relation $\mathbf{D} = \langle \mathbf{\Xi} \mathbf{\Xi}^T \rangle$. As the noise processes $\xi_s$, $\xi_x$, and $\xi_y$ are uncorrelated, the off-diagonal elements of the diffusion matrix $\mathbf{D}$ are zero. Here $T$ represents the transpose of a matrix. The analytical solution of Eq.~(\ref{eq5}) yields variance and covariance associated with different CFFL motifs (see Appendix).

We now focus on the statistical measures of the motifs under consideration, i.e., the variance of the output $\Sigma(y)$. Using the expressions of variance and covariance given in the Appendix (see Eqs.~(\ref{eqa1}-\ref{eqa6})) we have decomposed the expression for variance of Y \cite{Nandi2021},
\begin{equation}
\label{eq6}
\Sigma(y) =  \Sigma_i (y) + \Sigma_o (y) + \Sigma_t (y) +  \Sigma_{ct} (y),
\end{equation}

\noindent where,
\numparts
\begin{eqnarray}
 \nonumber \\
\Sigma_i (y) & = & \underbrace{\langle y \rangle}_{\rm intrinsic}, 
\label{eq6a} \\
\Sigma_o (y) & = & \underbrace{\frac{\langle s \rangle f_{y,s}^{\prime 2}}{\mu_y (\mu_s + \mu_y)}}_{\rm one-step}, 
\label{eq6b} \\
\Sigma_t (y) & = &
\underbrace{\frac{\langle x \rangle f_{y,x}^{\prime 2}}{\mu_y (\mu_x + \mu_y)}
+ \frac{\langle s \rangle (\mu_s + \mu_x + \mu_y)  f_{x,s}^{\prime 2} f_{y,x}^{\prime 2}}{\mu_x \mu_y (\mu_s + \mu_x) (\mu_x + \mu_y) (\mu_s + \mu_y)}}_{\rm two-step}, 
\label{eq6c} \\
\Sigma_{ct} (y) & = &
\underbrace{\frac{2 \langle s \rangle (\mu_s + \mu_x + \mu_y)  f_{x,s}^{\prime} f_{y,x}^{\prime} f_{y,s}^{\prime}}{\mu_y (\mu_s + \mu_x) (\mu_x + \mu_y) (\mu_s + \mu_y)}}_{\rm cross~term}.
\label{eq6d}
\end{eqnarray}
\label{eq6p}
\endnumparts

\noindent In Eq.~(\ref{eq6}), the first term $\Sigma_i (y)$ appears due to intrinsic fluctuations of Y. Intrinsic fluctuations can be attributed to the Poissonian dynamics of Y without the influence of upstream component S and X. For Poissonian dynamics, the variance becomes equal to the mean that we observe here, i.e., $\Sigma_i (y) = \langle y \rangle$.
The second term $\Sigma_o (y)$ is the contribution of the one-step pathway, direct interaction between S and Y. The direct interaction is either activation (S~$\rightarrow$~Y) or repression (S~ $\dashv$~Y) of Y by S (see Fig.~\ref{fig1}). The nature of activation or repression is given by $f_{y,s}^{\prime^2}$.
The third term $\Sigma_t (y)$ is due to a two-step pathway, the interaction between S and Y via X. The first component of $\Sigma_t (y)$ is pure X-mediated fluctuations of Y, which are either activating or repressing in nature (see Fig.~\ref{fig1} and Table~\ref{table1}). The second component is the contribution of fluctuations of S to Y via X. 
We note that these pathway-specific fluctuations do not correspond to the fluctuations due to pure one-step and two-step cascades. Fluctuations propagated through one-step and two-step pathways can be influenced by their complementary pathways i.e., by two-step and one-step paths, respectively.
The fourth term $\Sigma_{ct} (y)$ is the result of the integration of fluctuations from the one-step pathway and two-step pathway. The cross-term arises due to the synchronized activity of the one-step and two-step pathway \cite{Nandi2021,Bruggeman2009}. When one of the two pathways becomes inactive the cross-term vanishes. In absence of the cross interaction a CFFL breaks down into pure one-step and two-step pathways (see Fig.~\ref{fig1}). For a pure one-step pathway Eq.~(\ref{eq6}) reduces to
\begin{equation}
\label{eq6o}
\Sigma_{OSC} (y) =  \Sigma_i (y) + \Sigma_o (y).
\end{equation}

\noindent On a similar note for a pure two-step pathways the output variance is given by
\begin{equation}
\label{eq6t}
\Sigma_{TSC} (y) =  \Sigma_i (y) + \Sigma_t (y),
\end{equation}

\noindent The above analysis shows that pathway-based fluctuations decomposition not only provides a quantitative estimation of the total variance of a complex GTRN, e.g., FFL but provides quantification of fluctuations of its building blocks. We note that $\Sigma_o (y)$, $\Sigma_t (y)$ and $\Sigma_{ct} (y)$ together constitute the extrinsic fluctuations of Y in a FFL.

We now define CV for the output Y
\begin{equation}
\label{eq7}
\eta_y = \sqrt{\Sigma (y)}/\langle y \rangle.
\end{equation}

\noindent In the rest of our analysis, to measure the noise associated with the output Y we use the square of CV, i.e., $\eta_y^2~(= \Sigma (y)/\langle y \rangle^2)$. As a result Eq.~(\ref{eq6}) becomes
\begin{equation}
\label{eq6n}
\eta_y^2 = \eta_{y,i}^2 + \eta_{y,o}^2 + \eta_{y,t}^2 + \eta_{y,ct}^2.
\end{equation}

\noindent In the above expression $\eta_{y,i}^2$ is the intrinsic contribution to the total noise of Y, $\eta_y^2$, and $\eta_{y,o}^2$, $\eta_{y,t}^2$, and $\eta_{y,ct}^2$ are the extrinsic contributions. It is important to mention that the intrinsic component of $\eta_y^2$ is dependent on the mean level of Y whereas the extrinsic components are solely dependent on the mean level of the upstream variables S and X \cite{Hilfinger2011}. As we will observe in Sec.~\ref{sec3} the extrinsic noise components are dependent on the mean level of S and X.
Following the definition given in Eq.~(\ref{eq7}) we also have CV for pure one-step pathway and two-step pathway
\begin{eqnarray}
\label{eq6on}
\eta_{y,OSC}^2 & = & \eta_{y,i}^2 + \eta_{y,o}^2, \\
\label{eq6tn}
\eta_{y,TSC}^2 & = & \eta_{y,i}^2 + \eta_{y,t}^2.
\end{eqnarray}

\noindent On a similar note, we quantify the fluctuations associated with X as $\eta_x^2 = \Sigma (x)/\langle x \rangle^2$.

The calculation presented up to Eq.~(\ref{eq5}) is applicable for Hill coefficient $n \geqslant 1$. 
It is important to note that the Hill function is an approximation of detailed mass action kinetics of gene regulation mechanism. The detailed mass action kinetics of activation and repression involves binding and unbinding of transcription factors to the promoter site of the gene of interest. The binding and unbinding kinetics takes place on a faster time scale compared to the process of transcription and translation (see Table~2.2 of Ref.~\cite{Alon2006} and \cite{Govern2014}). Moreover, fluctuations associated with the binding and unbinding kinetics of the transcription factors do not appear in the Hill coefficient. Given this information linearization of the Hill function around steady state captures the essential features of gene regulation mechanism under certain parametric conditions.
In the next section, we show results for $n = 1, 2$. We exclude results for $n > 2$, say $n = 3$, etc., as the functions $f_x$ and $f_y$ are highly nonlinear in this regime. Results obtained using higher order nonlinear functions shows disagreement between theory and simulation \cite{Roy2021}.

The steady state population of the system variables are (see Eqs.(\ref{eq1}-\ref{eq3})
\begin{eqnarray}
\label{eq8}
\langle s \rangle & = & f_s(\langle s \rangle)/\mu_s, \langle x \rangle = f_x(\langle s \rangle, \langle x \rangle)/\mu_x,\nonumber \\
\langle y \rangle & = & f_y(\langle s \rangle, \langle x \rangle, \langle y \rangle)/\mu_y.
\end{eqnarray}

\noindent The mean copy numbers of $\langle x \rangle$ and $\langle y \rangle$ used are 100 and 100, respectively, with unit molecules/V. An Earlier study on signal propagation in biochemical motifs shows that separation of time scale is required for maximum fluctuations transmission \cite{Maity2015}. In the present work, we use $\mu_s \ll \mu_x \ll \mu_y$ so that X can sense the fluctuations of S and Y can sense fluctuations of X, effectively. In other words, the condition $\mu_s \ll \mu_x$ allows X to sense the fluctuations of S effectively as S fluctuates relatively on a slower time scale. For $\mu_s \gg \mu_x$, on the other hand, X would sense the mean level of S rather than its fluctuations \cite{Bruggeman2009}. In the same spirit, $\mu_x \ll \mu_y$ allows Y to sense the fluctuations of X effectively.

To this end we use $\mu_s=0.1$ min$^{-1}$, $\mu_x=1$ min$^{-1}$, $\mu_y=10$ min$^{-1}$. We set  $K_{xy} = \langle x\rangle$. We further use $K_{sx} = 50$ (molecules/V) and $K_{sy} = 50$ (molecules/V). We use $\langle s\rangle$ as the tuning parameter. 
The mean field kinetics of S is given by $ds/dt=k_s-\mu_s s$ (see Table~\ref{table1}) where $k_s$ is maximal promoter activity associated with S. For a fixed value of the parameter $\mu_s$ one can tune the promoter activity $k_s$ to produce different steady state levels of S, i.e., $\langle s\rangle$. In the present work, we tune the parameter $k_s$ and use the steady state average population of signal $\langle s\rangle$ as an independent variable to compute different statistical measures.
Given the population of $\langle s \rangle$, $\langle x \rangle$ and $\langle y \rangle$ and model  parameters one can evaluate the kinetic parameters $k_s$, $k_x$ and $k_y$  using Eq.~(\ref{eq8}) and Table~\ref{table1} for different CFFL constructs.
We note that the theoretical analysis presented is done using a steady state population of S, X, and Y as the steady state population of the proteins (transcription factors) is important for their optimal function \cite{Alon2006}. In the present work, we have used a fixed steady state population of Y of all four CFFLs. Moreover, most of the parameters are kept fixed for mathematically controlled comparison \cite{Savageau1976}.

The explicit form of $f_s$ given in Table~\ref{table1} suggests that $k_s = \mu_s \langle s \rangle$ and is applicable for all the CFFL constructs shown in Fig.~\ref{fig1} independent of their architecture. For C1-FFL with additive integration mechanism we have $f_x = k_x s^n/(K_{sx}^n + s^n)$ which when used in Eq.~(\ref{eq8}) yields $k_x = \mu_x \langle x \rangle ((K_{sx}^n + \langle s \rangle^n)/\langle s \rangle^n)$. Similarly, the value of the kinetic parameter $k_x$ can be evaluated for other CFFL constructs and is shown in Table~\ref{table2}. Using the same approach (i.e., using Table~\ref{table1} and Eq.~(\ref{eq8})), the expressions of $k_y$ for all CFFL constructs are evaluated and tabulated in Table~\ref{table2}.

The kinetics of fluctuations propagation in all the CFFL motifs are simulated using the stochastic simulation algorithm or Gillespie's algorithm \cite{Gillespie1976,Gillespie1977}. 
Statistical averaging is done using the steady state output of $10^6$ independent trajectories.


\section{Results and discussion}
\label{sec3}

As mentioned earlier, we consider here two types of integration mechanisms for fluctuations propagation in four different CFFLs, from signal S to target Y. The integration mechanisms are either additive or multiplicative in nature. The mathematical representation for these two types of signal integration mechanisms is shown in Table~\ref{table1} (see $f_y (s, x, y)$). In one integration mechanism, fluctuations of S and X are integrated additively into Y while considering activation and/or repression kinetics. In another integration mechanism, multiplicative integration takes place using the same system components. Using these integration mechanisms, we quantify statistical measures associated with output Y.


\begin{figure*}[!t]
\includegraphics[width=1.0\columnwidth,angle=0]{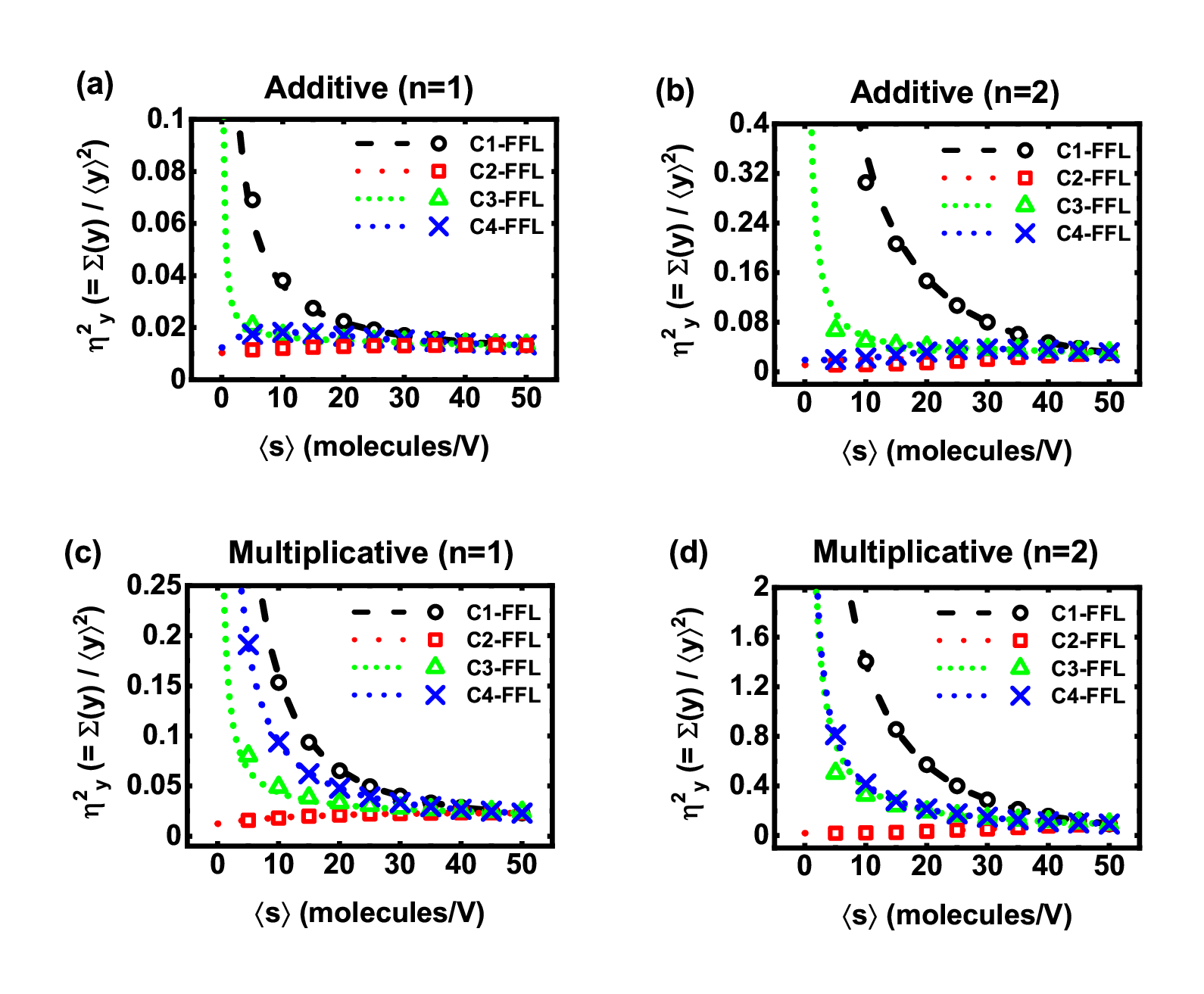}
\caption{Square of CV as a function of average signal strength $\langle s \rangle$ for different CFFL. Results for both additive and multiplicative integration mechanism are shown for $n = 1, 2$. The lines are due to theoretical calculation (Eqs.~(\ref{eq6n})) and symbols are generated using stochastic simulation algorithm \cite{Gillespie1976,Gillespie1977}. The parameters used are $\mu_s=0.1$ min$^{-1}$, $\mu_x=1$ min$^{-1}$, $\mu_y=10$ min$^{-1}$, $K_{xy} = 100$ (molecules/V), $K_{sx} = 50$ (molecules/V) and $K_{sy} = 50$ (molecules/V). The mean copy numbers of $\langle x \rangle$ and $\langle y \rangle$ are 100 and 100, respectively, with unit of molecules/V.
}
\label{fig3}
\end{figure*}

\subsection{C1-FFLs with multiplicative integration mechanism imposes maximum output fluctuations}

\begin{figure*}[!t]
\includegraphics[width=1.0\columnwidth,angle=0]{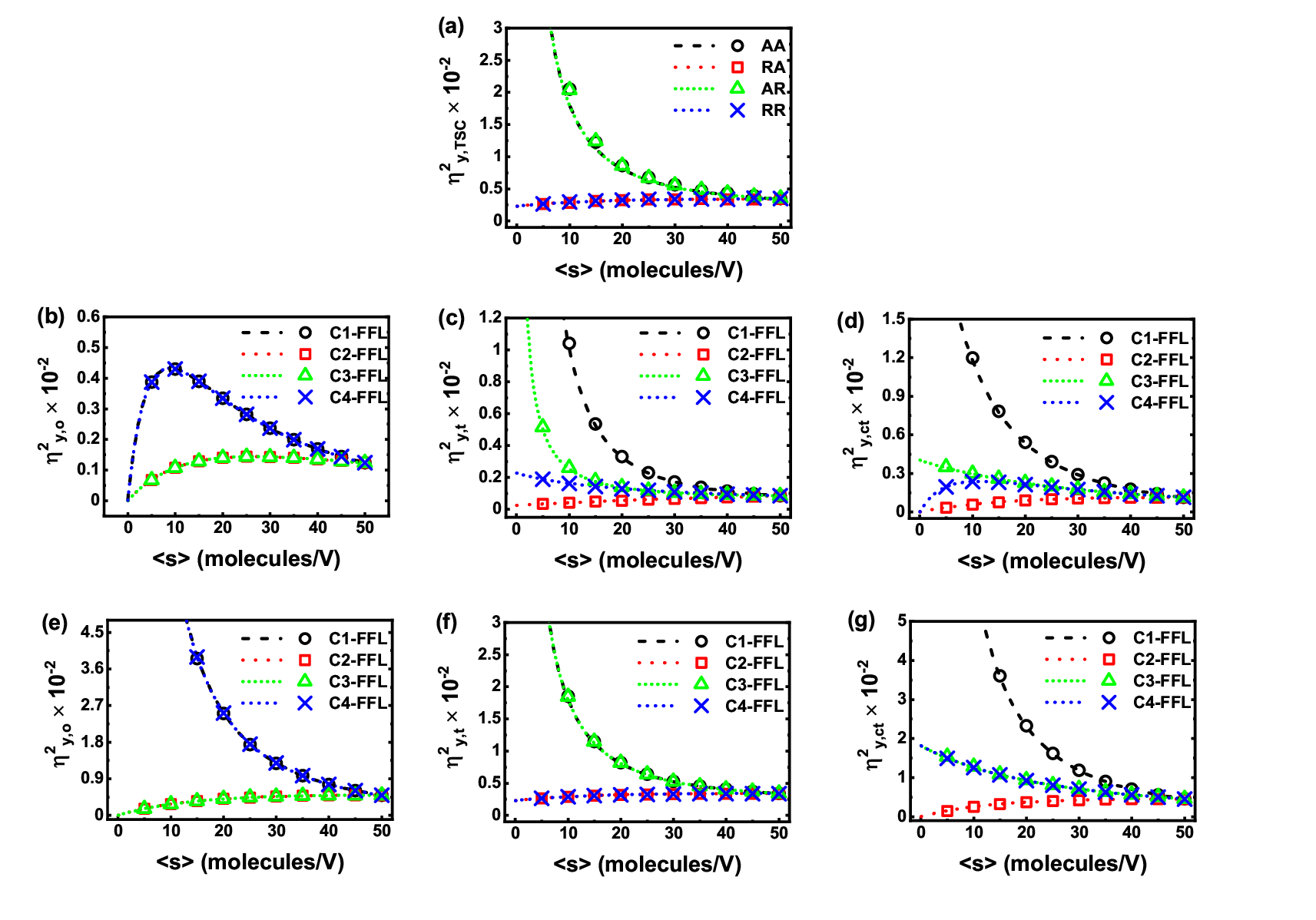}
\caption{(color online) Variation of decomposed fluctuations as a function of $\langle s\rangle$. (a) depicts coefficient of variation of Y $\eta_{y,TSC}^2$ of pure TSCs. (b) and (e) are one-step propagated fluctuations, $\eta_{y,o}^2$, for additive and multiplicative integration mechanisms, respectively. (c) and (f) are two-step propagated fluctuations $\eta_{y,t}^2$, for additive and multiplicative integration mechanisms, respectively. (d) and (g) are fluctuations due to cross term $\eta_{y,ct}^2$, for additive and multiplicative integration mechanisms, respectively. The lines are due to theoretical calculation (see Eq.~(\ref{eq6n})) and symbols are generated using stochastic simulation algorithm \cite{Gillespie1976,Gillespie1977}. The parameters used are $\mu_s=0.1$ min$^{-1}$, $\mu_x=1$ min$^{-1}$, $\mu_y=10$ min$^{-1}$, $K_{xy} = 100$ (molecules/V), $K_{sx} = 50$ (molecules/V) and $K_{sy} = 50$ (molecules/V). The mean copy numbers of $\langle x \rangle$ and $\langle y \rangle$ are 100 and 100, respectively, with unit of molecules/V.
}
\label{fig4}
\end{figure*}

Fig.~\ref{fig3} shows CV$^2$ of output Y, $\eta_y^2$ as a function of average signal strength $\langle s \rangle$. For both additive and multiplicative mechanisms, we observe decaying profiles of $\eta_y^2$. We note that the multiplicative integration mechanism generates a higher magnitude of fluctuations of Y (Fig.~\ref{fig3}(c,d)) in comparison to additive integration mechanisms (Fig.~\ref{fig3}(a,b)). Among the four CFFLs, C1-FFL, where all interactions are activating in nature, shows a high magnitude of fluctuations compared to the other three CFFLs irrespective of the integration mechanism. In other words, C1-FFL with a multiplicative integration mechanism exhibits maximum fluctuations of Y. Fig.~\ref{fig4}(a) shows only the CV$^2$ of Y for AA, RA, AR, and RR motifs where clustering of AA and AR (RA and RR) occurs. These four TSC motifs are building blocks of the four CFFLs. We anticipate that the trend of degeneracy has been lifted due to the addition of two types of the one-step pathway, viz., A and R to their respective TSC counterpart (see Fig.~\ref{fig1}). Fig.~\ref{fig3} shows the non-degeneracy in fluctuations of output Y. To understand this phenomena we now look at the individual components of $\eta_y^2$, i.e., $\eta_{y,o}^2$, $\eta_{y,t}^2$ and $\eta_{y,ct}^2$ while considering the contribution of additive and multiplicative integration mechanisms.

\subsection{Fluctuations decomposition gives rise to diverse flavours of degeneracy and non-degeneracy}


\subsubsection{Fluctuations propagation along one-step (direct) pathway}

The first term of Eq.~(\ref{eq6n}) stands for the intrinsic fluctuations of Y, $\eta_{y,i}^2$. The intrinsic fluctuations are the same for all four CFFLs irrespective of the types of integration mechanisms. Fluctuations propagation along the direct path (one-step), $\eta_{y,o}^2$ is plotted in Fig.~\ref{fig4}(b) and Fig.~\ref{fig4}(e) for additive and multiplicative mechanisms, respectively. Both the figures show a collating nature between C1 and C4-FFLs (C2 and C3-FFLs). The collating behavior can be understood from the architectural similarity in the direct pathways of CFFLs. In C1 and C4-FFLs, it is activating in nature whereas in C2 and C3-FFLs it is of a repressing type. It is important to mention that, the collating nature of $\eta_{y,o}^2$ is independent of the nature of integration mechanisms. However, the magnitude and nature of the profiles of $\eta_{y,o}^2$ for additive integration differ from its multiplicative counterpart. To analyze this behavior we look at the explicit expression of $\eta_{y,o}^2$ for four CFFLs. To this end, we use the expression of $f_{y,s}^{\prime 2}$ and $k_y$ from Table~\ref{table1} and Table~\ref{table2}, respectively. Further, we simplify the expressions of $\eta_{y,o}^2$ by using $K_{xy}=\langle x\rangle$. Finally for $n=1$ case we have the following expressions of $\eta_{y,o}^2$ as a function of $\langle s\rangle$,
\numparts
\begin{eqnarray}
\eta_{y,o}^2 & = & A^o \times \frac{\langle s\rangle}{(K_{sy} + \langle s \rangle)^2 (K_{sy}+3 \langle s\rangle)^2} \; ({\rm C1-FFL, \; additive}), 
\label{eq9a} \\
\eta_{y,o}^2 & = & A^o \times \frac{\langle s\rangle}{(K_{sy} + \langle s \rangle)^2 (3K_{sy}+\langle s\rangle)^2} \; ({\rm C2-FFL, \; additive}),
\label{eq9b} \\
\eta_{y,o}^2 & = & A^o \times \frac{\langle s\rangle}{(K_{sy} + \langle s \rangle)^2 (3K_{sy}+\langle s\rangle)^2} \; ({\rm C3-FFL, \; additive}), 
\label{eq9c} \\
\eta_{y,o}^2 & = & A^o \times \frac{\langle s\rangle}{(K_{sy} + \langle s \rangle)^2 (K_{sy}+3 \langle s\rangle)^2} \; ({\rm C4-FFL, \; additive}), 
\label{eq9d} \\
\eta_{y,o}^2 & = & \frac{A^o}{4} \times 
\frac{1}{\langle s\rangle (K_{sy}+\langle s\rangle)^2} 
\; ({\rm C1-FFL, \; multiplicative}), 
\label{eq9e} \\
\eta_{y,o}^2 & = & \frac{A^o}{4 K_{sy}^2} \times 
\frac{\langle s\rangle}{(K_{sy}+\langle s\rangle)^2}
\; ({\rm C2-FFL, \; multiplicative}), 
\label{eq9f} \\
\eta_{y,o}^2 & = & \frac{A^o}{4 K_{sy}^2} \times 
\frac{\langle s\rangle}{(K_{sy}+\langle s\rangle)^2} 
\; ({\rm C3-FFL, \; multiplicative}), 
\label{eq9g} \\
\eta_{y,o}^2 & = & \frac{A^o}{4} \times 
\frac{1}{\langle s\rangle (K_{sy}+\langle s\rangle)^2} 
\; ({\rm C4-FFL, \; multiplicative}),
\label{eq9h}
\end{eqnarray}
\endnumparts

\noindent where $A^o=4 \mu_y K_{sy}^4/(\mu_s+\mu_y)$. Eqs.~(\ref{eq9a}-\ref{eq9h}) shows that $\eta_{y,o}^2$ is the same for C1 and C4-FFLs (also for C2 and C3-FFLs) for both additive and multiplicative mechanisms. Moreover, it also explains the dissimilar magnitude of variations in $\eta_{y,o}^2$ between additive and multiplicative integrations. Hence, C1 and C4-FFLs (C2 and C3-FFLs) are degenerate with respect to the fluctuation propagation along the one-step pathway.

For additive integration mechanism, Eqs.~(\ref{eq9a}-\ref{eq9d}) suggests that at low $\langle s \rangle$, $\eta_{y,o}^2 \sim \langle s \rangle$ which takes into account of the initial growing nature of the profiles shown in Fig.~\ref{fig4}b. In the high $\langle s \rangle$ limit, $\eta_{y,o}^2 \sim 1/\langle s \rangle^3$ that takes care of the decaying nature of the profiles in Fig.~\ref{fig4}b. In the low $\langle s \rangle$ limit, a constant factor 1/9 appears in Eqs.~(\ref{eq9b}-\ref{eq9c}) which makes a difference in amplitude between the profiles of C1-FFL (C4-FFL) and C2-FFL (C3-FFL).
In the case of multiplicative integration mechanism, Eqs.~(\ref{eq9e},\ref{eq9h}) suggests that $\eta_{y,o}^2 \sim 1/\langle s \rangle$ and $\eta_{y,o}^2 \sim 1/\langle s \rangle^3$ for low and high $\langle s \rangle$, respectively, which explains the decaying nature of the profiles of C1- and C4-FFL in Fig.~\ref{fig4}e. On the other hand Eqs.~(\ref{eq9f}, \ref{eq9g}) suggest that $\eta_{y,o}^2 \sim \langle s \rangle$ for a wide range of $\langle s \rangle$, that explains the nature of profiles for C2- and C3-FFL in Fig.~\ref{fig4}e.

\subsubsection{Fluctuations propagation along two-step (indirect) pathway}

Fluctuations propagation along the two-step (indirect) pathway, $\eta_{y,t}^2$ is plotted  as a function of $\langle s\rangle$ in Fig.~\ref{fig4}(c) and Fig.~\ref{fig4}(f) for additive and multiplicative integration mechanisms, respectively. For additive case, $\eta_{y,t}^2$ is non-degenerative in nature for all the four CFFLs. On the other hand, the multiplicative integration mechanism brings in degeneracy in the $\eta_{y,t}^2$ profiles. We note that $\eta_{y,TSC}^2$ profiles for pure TSC motifs (Fig.~\ref{fig4}(a)) are exactly reproduced in $\eta_{y,t}^2$ profiles of two-step pathway with multiplicative integration mechanism. Here, in Fig.~\ref{fig4}(f), $\eta_{y,t}^2$ of C1- and C3-FFL overlap with each other as do C2- and C4-FFL. As mentioned earlier, similar collating nature occurs between AA and AR, and RA and RR motifs (see Fig.~\ref{fig4}(a)). From the profiles of Figs.~\ref{fig4}(a),~(c),~and~(f), one can infer that CFFLs with multiplicative integration mechanism shows degeneracy in $\eta_{y,t}^2$ similar to $\eta_{y,TSC}^2$ of pure TSC. In the multiplicative integration mechanism, input from both the direct and indirect pathways is necessary to build up the fluctuations space of Y. On the other hand, for the additive integration mechanism, one of the pathways (direct or indirect) is sufficient enough to generate the fluctuations space of target Y. Thus with the additive integration mechanism, the direct and indirect pathways act separately to generate a non-degenerate fluctuations space of target Y. Keeping the aforesaid discussion in mind, the similarity between Fig.~\ref{fig4}(a) and Fig.~\ref{fig4}(f) can be understood. In other words, although in CFFL with a multiplicative integration mechanism both the direct and indirect pathways are in action, due to the above reasoning these pathways act as an effective single pathway as happens in pure TSC.

To make a quantitative understanding of this phenomenon, we look at the expression of $\eta_{y,t}^2$ ($= \Sigma_t (y)/\langle y \rangle^2$) which involves two functions, $f_{x,s}^\prime$ and $f_{y,x}^\prime$. The other terms of $\eta_{y,t}^2$ are kept constant for all CFFLs during calculation. Using Table~\ref{table1}, we derive $f_{x,s}^\prime$ and $f_{y,x}^\prime$ which contain $k_x$ and $k_y$ whose simplified forms are given in Table~\ref{table2}. We further use $K_{yx}=\langle x\rangle$ into the expressions of $f_{x,s}^\prime$ and $f_{y,x}^\prime$ to finally derive the expressions of $\eta_{y,t}^2$ as a function of $\langle s\rangle$,
\numparts
\begin{eqnarray}
\eta_{y,t}^2 & = & \left(\frac{K_{sy}+\langle s\rangle}
{K_{sy}+3 \langle s\rangle} \right)^2 \times 
\left( A_1^t + A_2^t 
\frac{1}{\langle s\rangle (K_{sx}+\langle s\rangle)^2} \right)
\; ({\rm C1-FFL, \; additive}), 
\label{eq10a} \\
\eta_{y,t}^2 & = & \left(\frac{K_{sy}+\langle s\rangle}
{3K_{sy}+\langle s\rangle} \right)^2 \times 
\left( A_1^t + A_2^t 
\frac{\langle s\rangle}{K_{sx}^2 (K_{sx}+\langle s\rangle)^2} \right)
\; ({\rm C2-FFL, \; additive}), 
\label{eq10b} \\
\eta_{y,t}^2 & = & \left(\frac{K_{sy}+\langle s\rangle}
{3K_{sy}+\langle s\rangle} \right)^2 \times 
\left( A_1^t + A_2^t 
\frac{1}{\langle s\rangle (K_{sx}+\langle s\rangle)^2} \right)
\; ({\rm C3-FFL, \; additive}), 
\label{eq10c} \\
\eta_{y,t}^2 & = & \left(\frac{K_{sy}+\langle s\rangle}
{K_{sy}+3\langle s\rangle} \right)^2 \times 
\left( A_1^t + A_2^t 
\frac{\langle s\rangle}{K_{sx}^2 (K_{sx}+\langle s\rangle)^2} \right)
\; ({\rm C4-FFL, \; additive}),
\label{eq10d} \\
\eta_{y,t}^2 & = &  A_1^t + A_2^t
\frac{1}{\langle s\rangle (K_{sx}+\langle s\rangle)^2} 
\; ({\rm C1-FFL, \; multiplicative}), 
\label{eq10e} \\
\eta_{y,t}^2 & = &  A_1^t + A_2^t 
\frac{\langle s\rangle}{K_{sx}^2 (K_{sx}+\langle s\rangle)^2} 
\; ({\rm C2-FFL, \; multiplicative}), 
\label{eq10f} \\
\eta_{y,t}^2 & = & A_1^t + A_2^t 
\frac{1}{\langle s\rangle (K_{sx}+\langle s\rangle)^2}
\; ({\rm C3-FFL, \; multiplicative}), 
\label{eq10g} \\
\eta_{y,t}^2 & = & A_1^t + A_2^t 
\frac{\langle s\rangle}{K_{sx}^2 (K_{sx}+\langle s\rangle)^2} 
\; ({\rm C4-FFL, \; multiplicative}),
\label{eq10h}
\end{eqnarray}
\endnumparts

\noindent where $A_1^t=\mu_y /(4 \langle x\rangle (\mu_x+\mu_y))$ and $A_2^t=\mu_x \mu_y K_{sx}^2 (\mu_s+\mu_x+\mu_y)/(4 (\mu_s+\mu_x) (\mu_x+\mu_y) (\mu_s+\mu_y))$. Eqs.~(\ref{eq10a}-\ref{eq10h}) shows that $\eta_{y,t}^2$ of all the four CFFLs with additive integration mechanism are different from each other as observed in Fig.~\ref{fig4}(c). For the multiplicative integration mechanism, $\eta_{y,t}^2$ associated with C1- and C3-FFLs (C2- and C4-FFLs) are exactly similar which gets reflected in Fig.~\ref{fig4}(f).

For additive integration mechanism, Eqs.~(\ref{eq10a}, \ref{eq10c}) suggests that $\eta_{y,t}^2 \sim 1/\langle s \rangle$ and $\eta_{y,t}^2 \sim 1/\langle s \rangle^3$ for low and high values of $\langle s \rangle$, respectively, which explains decaying profiles of C1- and C3-FFL (starting from a high value) in Fig.~\ref{fig4}c. For C2- and C4-FFL, however, $\eta_{y,t}^2 \sim \langle s \rangle$ for a wide range of $\langle s \rangle$ that takes care of the nature of their profiles.
In the case of multiplicative integration mechanism, $\eta_{y,t}^2 \sim 1/\langle s \rangle$ and $\eta_{y,t}^2 \sim 1/\langle s \rangle^3$ for low and high values of $\langle s \rangle$, respectively (see Eqs.~(\ref{eq10e}, \ref{eq10g})), which takes care of the decaying profiles of C1- and C3-FFL in Fig.~\ref{fig4}f. On the other hand, for C2- and C4-FFL $\eta_{y,t}^2 \sim \langle s \rangle$ (see Eqs.~(\ref{eq10f}, \ref{eq10h})) for a wide range of $\langle s \rangle$ values.

\subsubsection{Fluctuations propagation due to cross term}

Fig.~\ref{fig4}(d) and Fig.~\ref{fig4}(g) shows $\eta_{y,ct}^2$, the fluctuations due to cross term, as a function of $\langle s\rangle$ for additive and multiplicative integration mechanisms, respectively. The cross term appears due to lateral overlap of one-step and two-step pathways forming FFLs (see Fig.~\ref{fig1}). The cross term vanishes if one of the pathways is turned off. The cross terms in the additive integration mechanism are non-degenerate in nature. For the multiplicative integration mechanism, the cross terms for C3- and C4-FFLs show degeneracy while the others (C1- and C2-FFLs) are non-degenerate in nature. The expression of $\Sigma_{ct}(y)$ given in Eq.~(\ref{eq6d}) contains $f_{x,s}^\prime$, $f_{y,x}^\prime$ and  $f_{y,s}^\prime$ which can be evaluated from Table~\ref{table1}. After substituting the expressions of $k_x$ and $k_y$ using Table~\ref{table2} in these three functions and using $K_{xy}=\langle x\rangle$, we have the following forms of $\eta_{y,ct}^2$,
\numparts
\begin{eqnarray}
\eta_{y,ct}^2 &=&
2 A^{ct} \times \frac{1}
{(K_{sx}+\langle s\rangle) (K_{sy} + 3\langle s\rangle)^2},
\; ({\rm C1-FFL, \; additive}) 
\label{eq12a} \\
\eta_{y,ct}^2 &=&
\frac{2 A^{ct}}{K_{sx}} \times 
\frac{\langle s\rangle}
{(K_{sx}+\langle s\rangle) (3K_{sy} + \langle s\rangle)^2},
\; ({\rm C2-FFL, \; additive})  
\label{eq12b} \\
\eta_{y,ct}^2 &=&
2 A^{ct} \times 
\frac{1}
{(K_{sx}+\langle s\rangle) (3K_{sy} + \langle s\rangle)^2},
\; ({\rm C3-FFL, \; additive}) 
\label{eq12c} \\
\eta_{y,ct}^2 &=&
\frac{2 A^{ct}}{K_{sx}} \times 
\frac{\langle s\rangle}
{(K_{sx}+\langle s\rangle) (K_{sy} +3 \langle s\rangle)^2},
\; ({\rm C4-FFL, \; additive}) 
\label{eq12d} \\
\eta_{y,ct}^2 &=&
A^{ct} \times  
\frac{1}
{\langle s\rangle (K_{sx}+\langle s\rangle) (K_{sy} + \langle s\rangle)},
\; ({\rm C1-FFL, \; multiplicative}) 
\label{eq12e} \\
\eta_{y,ct}^2 &=&
\frac{A^{ct}}{K_{sx} K_{sy}} \times 
\frac{\langle s\rangle}
{(K_{sx}+\langle s\rangle) (K_{sy} + \langle s\rangle)},
\; ({\rm C2-FFL, \; multiplicative}) 
\label{eq12f} \\
\eta_{y,ct}^2 &=&
\frac{A^{ct}}{K_{sy}} \times 
\frac{1}
{(K_{sx}+\langle s\rangle) (K_{sy} + \langle s\rangle)},
\; ({\rm C3-FFL, \; multiplicative}) 
\label{eq12g} \\
\eta_{y,ct}^2 &=&
\frac{A^{ct}}{K_{sx}} \times 
\frac{1}
{(K_{sx}+\langle s\rangle) (K_{sy} + \langle s\rangle)},
\; ({\rm C4-FFL, \; multiplicative}),
\label{eq12h}
\end{eqnarray}
\endnumparts

\noindent where, $A^{ct}=\mu_x \mu_y K_{sx} K_{sy} (\mu_s+\mu_x+\mu_y)/(\mu_s+\mu_x)(\mu_x+\mu_y)(\mu_s+\mu_y)$. Eq.~(\ref{eq12a}-\ref{eq12h}) shows that, all the $\eta_{y,ct}^2$ for CFFLs with additive integration mechanism are different. The $\eta_{y,ct}^2$ for C3- and C4-FFLs with multiplicative mechanism are similar in nature given $K_{sx}=K_{sy}=50$ molecules/V (see Sec.~\ref{sec2}).

For additive integration mechanism, Eqs.~(\ref{eq12a},\ref{eq12c}) shows that $\eta^2_{y,ct} \sim 1/\langle s \rangle^3$ for high $\langle s \rangle$ and explains the decaying nature of C1- and C3-FFL profiles in Fig.~\ref{fig4}d. In the low $\langle s \rangle$ limit, C3-FFL has an additional constant factor of 1/3 which reduces the initial value of the curve. For C2- and C4-FFL, however, $\eta^2_{y,ct} \sim \langle s \rangle$ for low $\langle s \rangle$ and $\eta^2_{y,ct} \sim 1/\langle s \rangle^2$ for high $\langle s \rangle$ (see Eqs.~(\ref{eq12b},\ref{eq12d})) which gets reflected in the initial growth and then subsequent decaying profile.
In multiplicative integration mechanism, $\eta^2_{y,ct} \sim 1/\langle s \rangle^3$ for C1-FFL (see Eq.~(\ref{eq12e})) that takes care of decaying profile shown in Fig.~\ref{fig4}g. The initial high value is due to the constant factor $A^{ct}$. For C3- and C4-FFL $\eta^2_{y,ct} \sim 1/\langle s \rangle^2$ with a common prefactor (as $K_{sy} = K_{sx}$) (see Eqs.~(\ref{eq12g},\ref{eq12h})). For C2-FFL, however, $\eta^2_{y,ct} \sim \langle s \rangle$ and $\eta^2_{y,ct} \sim 1/\langle s \rangle$ for low and high value of $\langle s \rangle$, respectively (see Eq.~(\ref{eq12f})).


\section{Conclusion}
\label{sec4}

In the present communication, we undertake four types of CFFLs with additive and multiplicative fluctuations integration mechanism at the output Y. Fig.~\ref{fig3} shows a non-degenerate nature of the CFFLs with respect to the fluctuations of the output Y. To analyze these results, we decomposed the output fluctuations into four major components, i.e., intrinsic, one-step, two-step, and cross term. In Fig.~\ref{fig4} we show the nature of each decomposed term as a function of $\langle s\rangle$. Mathematical analysis supplemented by numerical simulation reveals the following key results,

\begin{enumerate}

\item Fluctuations propagation along the one-step pathway, $\eta_{y,o}^2$, shows degeneracy in C1- and C4-FFLs, and C2- and C3-FFLs irrespective of integration mechanisms. Eq.~(\ref{eq9a}-\ref{eq9h}) explicitly shows such variations in $\eta_{y,o}^2$. Moreover, the architecture of one-step pathways in CFFLs also validates the presence of such degeneracy.

\item Fluctuations propagated through a two-step pathway, $\eta_{y,t}^2$, show degeneracy in C1- and C3-FFLs, and in C2- and C4-FFLs for multiplicative integration mechanism. On the other hand, such degeneracy is completely lifted for the additive integration mechanism. Eq.~(\ref{eq10a}-\ref{eq10d}) suggests that nature of non-degeneracy is observed due to different mathematical forms of $\eta_{y,t}^2$.

\item Analysis of the cross term reveals non-degeneracy for additive integration mechanism and partial degeneracy (C3 and C4-FFLs are degenerate, see Eq.~(\ref{eq12g}-\ref{eq12h})) for multiplicative integration mechanism. 
We note that the mathematical expressions in Eq.~(\ref{eq12g})~and~(\ref{eq12h}) differ due to the factors $1/K_{sy}^2$ and $1/K_{sx}^2$. In the present work, however, we use $K_{sy}=K_{sx}=50$ molecules/V. Due to this parametric dependency, we observe collating features in the cross-term of C3- and C4-FFL with a multiplicative integration scheme.
\end{enumerate}

Fig.~\ref{fig4} shows a diverse nature of fluctuations propagation in CFFLs with degenerate and non-degenerate characteristics. When the individual decomposed terms are combined to get $\eta_y^2$, an overall non-degeneracy is observed among the CFFLs with both additive and multiplicative integration mechanisms. Hence, the integration of different types of parallel pathways leads to the lifting of degeneracy in output fluctuations of CFFLs. In addition, we quantitatively explain the higher magnitude of fluctuations exhibited by C1-FFL with both additive and multiplicative integration mechanisms. Our study provides an intuitive insight into the propagation of the fluctuations along CFFLs in terms of pathway specific fluctuations.

Fluctuations in gene expression are sometimes detrimental for the cell, while they can be beneficial for cell fate and decision making \cite{Kaern2005,Eldar2010}. A biological motif with high output noise (fluctuations) leads to greater variability and is capable of dealing with diverse environmental cues. A specific architectural construct with high output noise might help in increasing the adaptive nature of the motif and make it an abundant candidate through the evolutionary selection process. The high abundance of C1-FFL could be attributed to its high noise profile which permits the motif to have greater diversity to cope with distinct environmental situations. Noise performances, thus, might play an important role in the selection of C1-FFL by evolution as the most abundant recurring network in GTRNs.

In this context, output fluctuations decomposition and degeneracy in fluctuations propagation help to understand the mechanism of fluctuations propagation along different sub-architectural pathways, which may provide an intuitive idea about the cellular decision-making governed by fluctuations level. Our analysis of fluctuations decomposition and finding degeneracy therein classify distinct pathways of fluctuations propagation into specific categories. Such categorical classification on the basis of fluctuations decomposition can be extended for several biologically relevant networks, e.g., protein-protein interaction network, signal transduction network, metabolic networks etc., which are ubiquitous in nature. In addition, pathway-based analysis might reveal novel features of signal propagation in IFFL as well as different feedback loop motifs \cite{Alon2006}. We aim to communicate work in this direction in the near future.


\ack

M.N. thanks SERB, India, for National Post-Doctoral Fellowship [PDF/2022/001807].


\appendix

\section*{Appendix: Solution of Lyapunov equation: Variance and covariance of system components}

\setcounter{section}{1}

Solution of Lyapunov equation (\ref{eq5}) provides analytical expressions of variance and covariance of system components within the purview of linearization. The different components of Lyapunov equation are,
\begin{eqnarray*}
\mathbf{J} = \left (
\begin{array}{ccc}
f^{\prime}_{s,s} (\langle s \rangle) - \mu_s & 0 & 0 \\
f^{\prime}_{x,s} (\langle s \rangle, \langle x \rangle) &
f^{\prime}_{x,x} (\langle s \rangle, \langle x \rangle) - \mu_x & 0 \\
f^{\prime}_{y,s} (\langle s \rangle, \langle x \rangle, \langle y \rangle) &
f^{\prime}_{y,x} (\langle s \rangle, \langle x \rangle, \langle y \rangle) &
f^{\prime}_{y,y} (\langle s \rangle, \langle x \rangle, \langle y \rangle) - \mu_y
\end{array}
\right ),
\end{eqnarray*}

\begin{eqnarray*}
\mathbf{\Sigma} = \left (
\begin{array}{ccc}
\Sigma (s) & \Sigma (s,x) & \Sigma (s,y) \\
\Sigma (s,x) & \Sigma (x) & \Sigma (x,y) \\
\Sigma (s,y) & \Sigma (x,y) & \Sigma (y)
\end{array}
\right ),
\end{eqnarray*}

\noindent and
\begin{eqnarray*}
\mathbf{D} = \left (
\begin{array}{ccc}
\langle \xi_s (t) \xi_s (t) \rangle & \langle \xi_s (t) \xi_x (t) \rangle & \langle \xi_s (t) \xi_y (t) \rangle \\
\langle \xi_s (t) \xi_x (t) \rangle & \langle \xi_x (t) \xi_x (t) \rangle & \langle \xi_x (t) \xi_y (t) \rangle \\
\langle \xi_s (t) \xi_y (t) \rangle & \langle \xi_x (t) \xi_y (t) \rangle & \langle \xi_y (t) \xi_y (t) \rangle
\end{array}
\right ).
\end{eqnarray*}

\noindent Now using the statistical properties of noise processes mentioned in Sec.~II and inserting $\mathbf{J}$, $\mathbf{\Sigma}$ and $\mathbf{D}$ into Lyapunov equation (\ref{eq5}) we obtain the following expressions of variance and co-variance,
\begin{eqnarray}
\label{eqa1}
\Sigma (s) & = & \langle s \rangle, \\
\label{eqa2}
\Sigma (s,x) & = & \frac{\langle s \rangle f_{x,s}^{\prime}}{\mu_s + \mu_x}, \\
\label{eqa3}
\Sigma (s,y) & = & \frac{\langle s \rangle f_{y,s}^{\prime}}{\mu_s + \mu_y}
+ 
\frac{\langle s \rangle f_{x,s}^{\prime} f_{y,x}^{\prime}}{
(\mu_s + \mu_x) (\mu_s + \mu_y)}, \\
\label{eqa4}
\Sigma (x) & = & \langle x \rangle +
\frac{\langle s \rangle f_{x,s}^{\prime 2}
}{
\mu_x (\mu_s + \mu_x)
}, \\
\label{eqa5}
\Sigma (x,y) & = & \frac{\langle x \rangle f_{y,x}^{\prime}
}{
\mu_x + \mu_y
}
+ 
\frac{\langle s \rangle f_{x,s}^{\prime 2} f_{y,x}^{\prime} (\mu_s + \mu_x + \mu_y)
}{
\mu_x (\mu_s + \mu_x) (\mu_s + \mu_y) (\mu_x + \mu_y)
}
\nonumber \\
&& +
\frac{\langle s \rangle f_{x,s}^{\prime} f_{y,s}^{\prime} (2\mu_s + \mu_x + \mu_y)
}{
(\mu_s + \mu_x) (\mu_s + \mu_y) (\mu_x + \mu_y)
}, \\
\label{eqa6}
\Sigma (y) & = & \underbrace{\langle y \rangle}_{\rm intrinsic} 
+
\underbrace{\frac{\langle s \rangle f_{y,s}^{\prime 2}}{\mu_y (\mu_s + \mu_y)}}_{\rm one-step}
\nonumber \\
&& +
\underbrace{\frac{\langle x \rangle f_{y,x}^{\prime 2}}{\mu_y (\mu_x + \mu_y)}
+ 
\frac{\langle s \rangle (\mu_s + \mu_x + \mu_y)  f_{x,s}^{\prime 2} f_{y,x}^{\prime 2}}{\mu_x \mu_y (\mu_s + \mu_x) (\mu_x + \mu_y) (\mu_s + \mu_y)}}_{\rm  two-step}
\nonumber \\
& & +
\underbrace{\frac{2 \langle s \rangle (\mu_s + \mu_x + \mu_y)  f_{x,s}^{\prime} f_{y,x}^{\prime} f_{y,s}^{\prime}}{\mu_y (\mu_s + \mu_x) (\mu_x + \mu_y) (\mu_s + \mu_y)}}_{\rm cross~term}, \nonumber \\
& = & \Sigma_i (y) + \Sigma_o (y) + \Sigma_t (y) + \Sigma_{ct} (y) ,
\end{eqnarray}

\noindent where we have used $f_{s,s}^\prime=f_{x,x}^\prime=f_{y,y}^\prime=0$. This equality can easily be understood from Table~\ref{table1}. In Eq.~(\ref{eqa6}), the first term denotes intrinsic fluctuation of Y and denoted as $\Sigma_i (y)$. The second term considers the fluctuations due to direct interaction between S and Y and is denoted as $\Sigma_o (y)$ (one-step contribution). The third term is the contribution in fluctuations due to an indirect/two-step pathway and denoted as $\Sigma_t(y)$. The fourth contribution term appears due to the integration of the direct and indirect pathways forming FFLs and is denoted as $\Sigma_{ct}(y)$ (cross term).


\section*{References} 


\begin{thebibliography}{99}
\expandafter\ifx\csname url\endcsname\relax
  \def\url#1{{\tt #1}}\fi
\expandafter\ifx\csname urlprefix\endcsname\relax\def\urlprefix{URL }\fi
\providecommand{\eprint}[2][]{\url{#2}}

\bibitem{Milo2002}
Milo R, Shen-Orr S, Itzkovitz S, Kashtan N, Chklovskii D and Alon U 2002 {\em
  Science\/} {\bf 298} 824--827

\bibitem{Alon2006}
Alon U 2006 {\em An Introduction to Systems Biology: Design Principles of
  Biological Circuits\/} (CRC Press, Boca Raton)

\bibitem{Kaern2005}
Kaern M, Elston T~C, Blake W~J and Collins J~J 2005 {\em Nat. Rev. Genet.\/}
  {\bf 6} 451--464

\bibitem{Eldar2010}
Eldar A and Elowitz M~B 2010 {\em Nature\/} {\bf 467} 167--173

\bibitem{Shen-Orr2002}
Shen-Orr S~S, Milo R, Mangan S and Alon U 2002 {\em Nat. Genet.\/} {\bf 31}
  64--68

\bibitem{Mangan2003}
Mangan S and Alon U 2003 {\em Proc. Natl. Acad. Sci. U. S. A.\/} {\bf 100}
  11980--11985

\bibitem{Ma2004}
Ma H~W, Kumar B, Ditges U, Gunzer F, Buer J and Zeng A~P 2004 {\em Nucleic
  Acids Res.\/} {\bf 32} 6643--6649

\bibitem{Mangan2006}
Mangan S, Itzkovitz S, Zaslaver A and Alon U 2006 {\em J. Mol. Biol.\/} {\bf
  356} 1073--1081

\bibitem{Kalir2005}
Kalir S, Mangan S and Alon U 2005 {\em Mol. Syst. Biol.\/} {\bf 1} 2005.0006

\bibitem{Amit2007}
Amit I, Citri A, Shay T, Lu Y, Katz M, Zhang F, Tarcic G, Siwak D, Lahad J,
  Jacob-Hirsch J, Amariglio N, Vaisman N, Segal E, Rechavi G, Alon U, Mills
  G~B, Domany E and Yarden Y 2007 {\em Nat. Genet.\/} {\bf 39} 503--512

\bibitem{Kaplan2008}
Kaplan S, Bren A, Dekel E and Alon U 2008 {\em Mol. Syst. Biol.\/} {\bf 4} 203

\bibitem{Bleris2011}
Bleris L, Xie Z, Glass D, Adadey A, Sontag E and B Y 2011 {\em Mol. Sys.
  Biol.\/} {\bf 7} 519

\bibitem{Chen2013}
Chen S~H, Masuno K, Cooper S~B and Yamamoto K~R 2013 {\em Proc. Natl. Acad.
  Sci. U. S. A.\/} {\bf 110} 1964--1969

\bibitem{Sasse2015}
Sasse S~K and Gerber A~N 2015 {\em Pharmacol. Ther.\/} {\bf 145} 85--91

\bibitem{Alon2007}
Alon U 2007 {\em Nat. Rev. Genet.\/} {\bf 8} 450--461

\bibitem{Li2009}
Li X, Cassidy J~J, Reinke C~A, Fischboeck S and Carthew R~W 2009 {\em Cell\/}
  {\bf 137} 273--282

\bibitem{Herranz2010}
Herranz H and Cohen S~M 2010 {\em Genes Dev.\/} {\bf 24} 1339--1344

\bibitem{Ghosh2005}
Ghosh B, Karmakar R and Bose I 2005 {\em Phys. Biol.\/} {\bf 2} 36--45

\bibitem{Kittisopikul2010}
Kittisopikul M and Suel G~M 2010 {\em Proc. Natl. Acad. Sci. U.S.A.\/} {\bf
  107} 13300--13305

\bibitem{Maity2015}
Maity A~K, Chaudhury P and Banik S~K 2015 {\em PLoS ONE\/} {\bf 10} e0123242

\bibitem{Momin2020a}
Momin M~S~A, Biswas A and Banik S~K 2020 {\em Phys. Rev. E\/} {\bf 101} 022407

\bibitem{Biswas2022}
Biswas A 2022 {\em Phys. Rev. E\/} {\bf 105} 034406

\bibitem{Dunlop2008}
Dunlop M~J, Cox R~S, Levine J~H, Murray R~M and Elowitz M~B 2008 {\em Nat.
  Genet.\/} {\bf 40} 1493--1498

\bibitem{Momin2020b}
Momin M~S~A and Biswas A 2020 {\em Phys. Rev. E\/} {\bf 101} 052411

\bibitem{Nandi2021}
Nandi M 2021 {\em Theo. Biosci.\/} {\bf 140} 139--155

\bibitem{Roy2021}
Roy T~S, Nandi M, Biswas A, Chaudhury P and Banik S~K 2021 {\em Theo.
  Biosci.\/} {\bf 140} 295--306

\bibitem{Paulsson2004}
Paulsson J 2004 {\em Nature\/} {\bf 427} 415--418

\bibitem{Swain2002}
Swain P~S, Elowitz M~B and Siggia E~D 2002 {\em Proc. Natl. Acad. Sci. U. S.
  A.\/} {\bf 99} 12795--12800

\bibitem{Bintu2005}
Bintu L, Buchler N~E, Garcia H~G, Gerland U, Hwa T, Kondev J and Phillips R
  2005 {\em Curr. Opin. Genet. Dev.\/} {\bf 15} 116--124

\bibitem{Ziv2007}
Ziv E, Nemenman I and Wiggins C~H 2007 {\em PLoS ONE\/} {\bf 2} e1077

\bibitem{Tkacik2008a}
Tka{\v c}ik G, Callan C~G and Bialek W 2008 {\em Proc. Natl. Acad. Sci.
  U.S.A.\/} {\bf 105} 12265--12270

\bibitem{Tkacik2008b}
Tka{\v c}ik G, Gregor T and Bialek W 2008 {\em PLoS ONE\/} {\bf 3} e2774

\bibitem{Ronde2012}
de~Ronde W~H, Tostevin F and ten Wolde P~R 2012 {\em Phys. Rev. E\/} {\bf 86}
  021913

\bibitem{Elf2003}
Elf J and Ehrenberg M 2003 {\em Genome Res.\/} {\bf 13} 2475--2484

\bibitem{Swain2004}
Swain P~S 2004 {\em J. Mol. Biol.\/} {\bf 344} 965--976

\bibitem{Tanase2006}
T{\u a}nase-Nicola S, Warren P~B and ten Wolde P~R 2006 {\em Phys. Rev.
  Lett.\/} {\bf 97} 068102

\bibitem{Warren2006}
Warren P~B, T{\u a}nase-Nicola S and ten Wolde P~R 2006 {\em J. Chem. Phys.\/}
  {\bf 125} 144904

\bibitem{Kampen2007}
van Kampen N~G 2007 {\em Stochastic Processes in Physics and Chemistry, 3rd
  ed.\/} (North-Holland, Amsterdam)

\bibitem{Mehta2008}
Mehta P, Goyal S and Wingreen N~S 2008 {\em Mol. Syst. Biol.\/} {\bf 4} 221

\bibitem{Ronde2010}
de~Ronde W~H, Tostevin F and ten Wolde P~R 2010 {\em Phys. Rev. E\/} {\bf 82}
  031914

\bibitem{Swain2016}
Swain P~S 2016 {\em arXiv: q-bio.QM/1607.07806\/}

\bibitem{Elowitz2002}
Elowitz M~B, Levine A~J, Siggia E~D and Swain P~S 2002 {\em Science\/} {\bf
  297} 1183--1186

\bibitem{Raser2004}
Raser J~M and O'Shea E~K 2004 {\em Science\/} {\bf 304} 1811--1814

\bibitem{Keizer1987}
Keizer J 1987 {\em Statistical Thermodynamics of Nonequilibrium Processes\/}
  (Springer-Verlag, Berlin)

\bibitem{Paulsson2005}
Paulsson J 2005 {\em Phys. Life Rev.\/} {\bf 2} 157--175

\bibitem{Bruggeman2009}
Bruggeman F~J, Bl{\"u}thgen N and Westerhoff H~V 2009 {\em PLoS Comput.
  Biol.\/} {\bf 5} e1000506

\bibitem{Hilfinger2011}
Hilfinger A and Paulsson J 2011 {\em Proc. Natl. Acad. Sci. U. S. A.\/} {\bf
  108} 12167--12172

\bibitem{Govern2014}
Govern C~C and ten Wolde P~R 2014 {\em Proc. Natl. Acad. Sci. U. S. A.\/} {\bf
  111} 17486--17491

\bibitem{Savageau1976}
Savageau M~A 1976 {\em Biochemical Systems Analysis: A Study of Function and
  Design in Molecular Biology\/} (Addison-Wesley, Reading, MA)

\bibitem{Gillespie1976}
Gillespie D~T 1976 {\em J. Comp. Phys.\/} {\bf 22} 403--434

\bibitem{Gillespie1977}
Gillespie D~T 1977 {\em J. Phys. Chem.\/} {\bf 81} 2340--2361

\end{thebibliography}

\providecommand{\newblock}{}

\end{document}